# Towards Enhanced Usability of IT Security Mechanisms

How to Design Usable IT Security Mechanisms Using the Example of Email Encryption

Hans-Joachim Hof
Munich IT Security Research Group (MuSe)
Department of Computer Science and Mathematics
Munich University of Applied Sciences,
Lothstraße 64, 80335 Munich, Germany
email: hof@hm.edu

*Abstract*—Nowadays, advanced security mechanisms exist to protect data, systems, and networks. Most of these mechanisms are effective, and security experts can handle them to achieve a sufficient level of security for any given system. However, most of these systems have not been designed with focus on good usability for the average end user. Today, the average end user often struggles with understanding and using security mechanisms. Other security mechanisms are simply annoying for end users. As the overall security of any system is only as strong as the weakest link in this system, bad usability of IT security mechanisms may result in operating errors, resulting in insecure systems. Buying decisions of end users may be affected by the usability of security mechanisms. Hence, software providers may decide to better have no security mechanism then one with a bad usability. Usability of IT security mechanisms is one of the most underestimated properties of applications and systems. Even IT security itself is often only an afterthought. Hence, usability of security mechanisms is often the afterthought of an afterthought. This paper presents some guidelines that should help software developers to improve end user usability of security-related mechanisms, and analyzes common applications based on these guidelines. Based on these guidelines, the usability of email encryption is analyzed and an email encryption solution with increased usability is presented. The approach is based on an automated key and trust management. The compliance of the proposed email encryption solution with the presented guidelines for usable security mechanisms is evaluated.

*Keywords-usability; IT security; usable security; email encryption.*

## I. INTRODUCTION

This paper is an extension of the usability design guide presented in [1].

Any improvement of the overall security level of any system requires to improve the security level of all subsystems and available mechanisms as the overall security level of a system is determined by the weakest link in this system [2,3]. Howe et al. found that current software and approaches for security are not adequate for end users, because these mechanisms are missing ease of use [4]. In [2,3] end user are identified as weakest link in a company. Hence, improving the usability of security mechanisms helps to improve the overall security level of a system.

Examples of bad usability of security mechanisms are all around, some are discussed in Section IV. Bad usability of security mechanisms may slow down the adoption of a security system. This happened for example with email encryption. Today, it is very unlikely that an average user uses email encryption. Major problems for average users are key exchange and trust management, both having a very bad usability in common email encryption solutions. Figure 1 shows a completely useless error message during the generation of a key pair for email encryption in GPGMail [5], as one example of bad usability.

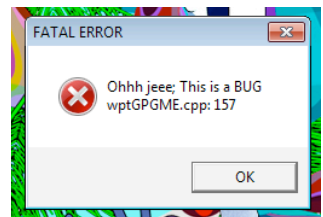

Figure 1. Error message during generation of a key pair for email encryption in GPGMail [5]

The use of email encryption in companies shows that an improved usability may lead to the adoption of the formerly despised technology. In companies, key exchange and trust management are usually not done by the users themselves, but they can rely on central infrastructures such as a central company directory with keys that are trusted by default (all employees). Such a directory ensures average users can use email encryption.

The example of email encryption shows that designing security mechanisms with good usability is worth an effort. For the ordinary software developer, i.e., non security expert, it makes sense not to implement core security mechanisms like encryption algorithms or signature algorithms. Those mechanisms are usually available in security libraries written by security experts and could be easily used by software developers. However, software developers often decide on how security mechanisms are integrated into an application. For example, when implementing an email encryption security solution like GPGMail [5], the software developer decides on the interfaces for setting up trust and importing keys. Both mechanisms are application specific, hence must be implemented by the application developers. Usually, these functionalities are exposed to the users, hence should have a good usability. Guidelines for usability that focus on IT security mechanisms and their integration into applications may help software developers to improve the usability of IT security mechanisms in these applications. This paper presents some guidelines that should help software developers to improve end user usability of security-related mechanisms. To underline the importance of the presented guidelines, weaknesses of security mechanisms in common applications regarding usability for end users are shown in an analysis of common applications and security mechanisms on basis of the presented guidelines. Other important aspects of software security, e.g., secure coding guidelines, testing of security, and threat analysis are out of scope of this paper.

The rest of this paper is structured as follows: Section II gives an overview on related work, especially on existing guidelines for usability. Section III presents guidelines for usable IT security mechanisms. Section IV analyzes the usability of some common security mechanisms and applications on the basis of the guidelines of Section III. Section V uses email encryption as an example on how to apply the guidelines on a problem from the field. An email encryption solution with good usability is presented. Section VI evaluates the usability of the email encryption solution on the basis of the guidelines presented in Section III. Section VII concludes the paper and gives an outlook on future work.

## II. RELATED WORK

Several standards focusing on usability in general exist, e.g., EN ISO 9241 [6]. In EN ISO 9241-11, which is part of EN ISO 9241, requirements for the usability of system are described. These requirements include effectiveness, efficiency and satisfaction. EN ISO 9241-10, another part of EN ISO 9241, lists requirements for usable user dialogs. However, the rules and guidelines of EN SIO 9241 are very general and not targeted on security mechanisms. The design guidelines presented in this paper interpret the general requirements and rules of EN ISO 9241 and its parts for the special case of security mechanisms. As the guidelines presented in this paper are focused only on the topic IT security, the presented guidelines are more detailed and may be easier to follow for software developers.

Other publications like [7-11] focus on the usability of security mechanisms in special applications (e.g., email encryption), or focus on the usability of special security mechanisms (e.g., use of passwords). The guidelines presented in this paper are more general such that they are useful for the design of a wide variety of applications and security mechanisms.

Existing guidelines for usability of security mechanisms like those in [12, 13] focus very much on user interface design. The design guide presented in this paper take a slightly different approach by focusing more on the security mechanism itself. It is considered possible to change the design of a security mechanism for the sake of good usability.

Markotten shows how to integrate user-centred security engineering into different phases of the software development process [14]. However, the emphasize of Markotten's work is more on integration of usability engineering into the software development process than on a design guide.

Several works on zero-configuration IT security exist, e.g., [15-22]. While zero-configuration can significantly improve the usability of an application, a systematic approach to usability for IT security is still missing. Zero-configuration may be one building block of usable security.

To summarize, previous works either are not focused on usability of IT security at all, are focused on one special aspect of usable IT security, or are focused on user interface design. This paper presents some guidelines for software developers to help them improve the usability of security-related functionality.

## III. GUIDELINES FOR GOOD USABILITY OF SECURITY MECHANISMS

The guidelines presented in this section are the result of several years in teaching IT security to beginners (and seeing their difficulties) as well as industrial experience in the design of products requiring IT security mechanisms that are operated by end users. The guidelines reflect our viewpoint on usability of security mechanisms. It is not assumed that those guidelines are complete. It is important to notice that the usability of any system depends on the specific user and his experiences, knowledge and context of use, which includes the task at hand, the equipment at hand, and the physical and social environment of the user. Hence, it is hard to objectively evaluate the usability of a system. However, we hope that the following set of nine design guidelines coming from the field may be of help for software developers:

**G1 Understandability, open for all users:** This paper focuses on usability for end users. The average end users should be able to use the security mechanism. Otherwise, the security mechanism is not useful for the intended audience. The average user neither has a special interest in IT security nor understands IT security. It is the responsibility of the software developer to hide as many security mechanisms as possible from the user. For those security mechanisms that are exposed to the end user it is necessary to get security awareness. The process of educating people is easier if suitable metaphors are used. A good metaphor is taken from everyday life of the average user and is easy to grasp. A good metaphor is simple but powerful in its meaning. Example: an email encryption application should not use the term "encrypted email." It is better to talk about a "secret message for xy" or "email readable only by xy" where xy is the receiver of the message.

Usable security should be available for all users. It should especially not discriminate against any group of people. For example, usable security mechanisms should not exclude disabled people that use special tools to access applications (e.g., Braille reader for vision impaired people). Example of compliance with G1: if captchas are used in an application, multiple versions of the captcha should exist. Each version of the captcha should address another sense.

**G2 Empowered users**: Ideally, a usable security mechanism should not be used to restrict the user in what he is doing or what he wants to do. This allows end users to efficiently fulfill their tasks. Efficiency is one of the general usability requirements of EN ISO 9241 [6]. The absence of user restrictions often results in a better acceptance of securi-ty by users. The focus of a security mechanism should be on protecting the user. Any security-motivated restriction of the user should be carefully evaluated regarding necessity for system security and adequateness. The user should at least have the impression that he is in control of the system and not the system is controlling him. Security mechanisms should interfere with the usual flow of user activities in the least possible way. Security mechanisms should allow the user to execute activities in any way he wants. Other drivers than protecting the user and the system should not be motivation for restrictions. Especially, users should not be restricted by a security mechanism for the only reason of copyright protection or other business reasons. While such security mechanisms are of great use for businesses, they constantly restrict the user, hence force him to bypass security mechanisms. As users are very imaginative in bypassing unwanted restrictions, it is very likely that a non-security-motivated restriction decreases the security level of a system. The Apple iPhone is a good example: as the phone enforces many restrictions, many user bypass the security mechanisms by using a jailbreak software to revoke those restrictions.

Another important rule is that the user should decide on trust relations. A security mechanism should not enforce trust relations given by a software vendor. The user should always have the possibility to revoke preinstalled trust relations. Trust relations should only be established in advance for the purpose of IT security. For example, having a preinstalled certificate to verify software patches is OK. Establishing trust relations out of business purposes should be avoided. Example of compliance with G2: applications should have an interface that lists preinstalled certificates. The user should have the possibility to revoke certificates and install custom certificates.

**G3 No jumping through hoops:** Users should only be forced to execute as little tasks as possible that exist only for IT security reasons. Otherwise, users get annoyed and refuse collaboration with IT security mechanisms. The ideal security mechanism does not interfere with user tasks at any time (also see G2) if it is not absolutely necessary to maintain the user's security.

An example on how to not design security mechanisms are captchas: the user is forced to read a nearly unreadable and meaningless combination of letters and numbers and enter it before he can execute the wanted task. Example of compliance with G3: an application that uses a challenge-response mechanism similar to hashcash [23] instead of a captcha to avoid abuse of a service by automated scripts.

**G4 Efficient use of user attention and memorization capability**: Users have problems memorizing data that does not belong to their social background. Hence, they tend to use all kind of optimization to reduce the amount of data they have to remember. This is why users only use few passwords for all logins where they need passwords. In [24], it is stated that an average user uses only 6.5 passwords for all his web accounts. A later survey [25] found that more than 80% of all participants of the survey reuse a set of password in different places. 73% of the participants use one password with slight modification on different accounts.

But not only does an average user use the same password more than once, he also selects easy to remember passwords as he is not good in memorizing passwords with a mix of upper and lower case letters, numbers and special characters. Hence, security mechanisms should require the user only to remember little data or no data at all. Example of compliance with G4: an application uses an existing account from another site for login, e.g., by using OpenID [26]. The user can use an existing account, hence does not have to remember another password.

Security mechanisms should only require as little interaction with the user as possible. The security mechanism should only requests the attention of the user if it is absolutely necessary. Interaction with the user should be done in the most minimalistic way. See also G1 for user interaction. Example of compliance with G4: an email encryption application that does not ask a user for each mail if he wants to encrypt the mail or not. Instead, the email application offers a configuration option to always encrypt mails. Additionally, the email composition window clearly states the current protection status and offers a possibility to override the preferences.

**G5 Only informed decisions:** A user only feels secure and cooperates with a system if the system does not ask too much of him. Hence, users should only have to make decisions they can decide on. If there is an important security decision to take, it must be ensured that the user has the capability to make this decision. This means that the user has enough information about the situation that requires him to make a decision, and it must be ensured that the average user is capable to make an informed decision on this issue. If it is not clear if the user can decide on an issue, the decision should be avoided. G5 is hard to achieve and requires a careful examination during the design of an application. Example of compliance with G5: an application automatically deals with unknown certificates and does not prompt a user for a decision (see Section IV.D).

**G6 Security as default:** Good usability requires efficiency. Hence, the user should not have to configure security when he first starts an application. Software for end users should always come preconfigured such that the software is reasonable secure and usable. All security mechanisms of a system should be delivered to the end user with a configuration that offers adequate security for the end users. If a preconfiguration is not possible, the configuration effort must be minimized for users. This requires an analysis of the security requirements of average users during software development prior to the deployment of the software to find the adequate security level for most users. Example of compliance with G6: a home wifi access point comes preconfigured with a random WiFi password.

**G7 Fearless System:** The security system should support a positive attitude of the user towards the security system. A user with a positive attitude towards security mechanisms is cooperative and more likely to not feel interrupted by security mechanisms. Hence, security mechanisms should protect the overall system in a way that the user neither has fear when the system is in a secure state nor feels secure when the system is not in a secure state. The security state of the sys-tem should be visible at all times. A security mechanism should be consistent in its communication with its user. A security mechanism should not use fear to force users to obey security policies or get a wanted reaction. G7 is hard to achieve and requires a careful examination during the design of an application.

**G8 Security guidance, educating reaction on user errors:** Users tend to make mistakes, especially in respect to IT security. It is important that the security system hinders the user to make mistakes. However, as blocked operations can be very frustrating for users, the response of the security system must provide information why a given operation was blocked and should also offer a solution on how the user could proceed. The solution must be adapted on the situation and should keep the overall security of the system in mind. A security system should guide the user in the usage of security mechanisms. Errors should be prevented and there should be ways to "heal" errors. Example of compliance with G8: when an email encryption application fails to encrypt an email because of a missing public key of the recipient, the error message should explain how to import certificates from and how to verify certificates by comparing fingerprints of keys. To "heal" the error, the email encryption application offers to send the mail as password-protected PDF and instruct the user to call the recipient and tell him the password for the PDF.

**G9 Consistency:** Consistency allows users to efficiently fulfill their tasks. Security mechanisms should fit into both the application and the system context where they are used. Security mechanisms should have the look and feel the user is used to. G9 is hard to achieve and requires a careful examination during the design of an application.

IV. ANALYSIS OF THE USABILITY OF COMMON SECURITY MECHANISMS AND APPLICATIONS ON BASIS OF THE PRESENTED GUIDELINES

In this section, common applications and security mechanisms are analyzed on basis of the guidelines given in Section III. The analysis identifies room for improvement in these applications and security mechanisms. It also shows some good examples for certain aspects of security usability.

*A. Email Encryption using GPGMail*

GPGMail is a popular open source email encryption solution for Mac home users [5]. The encryption process itself is fairly easy, usually requiring one click to enable email encryption. However, key and trust management requires significant effort. For a secure exchange of public keys, the user has to get the public key itself (e.g., from a key server or the homepage of the receiver of a message) and verify the authenticity of the key. Certificates may be in use. The authentication requires the use of another channel to communicate with the key owner (e.g., telephone or in person) and to read a number to the owner that is meaningless for the user. There is no guidance for this process. Then, the user has to change the trust of the exchanged public key. It gets more complicated when using a web of trust for trust management: for the web of trust to work, the user must decide on how trustworthy a person is to verify public keys/certificates in addition to managing direct trust into keys. The distinction between those different types of trust is very hard to understand for average users.

This application is compliant with the following guidelines:
- G2 (user decides on trust relations)
- G4 (minimal interaction)
- G7 (does not frighten user)
- G9 (usually good integration, depends on system, mail client)

This application is not compliant with the following guidelines:
- G1 (hard to understand trust management and process of key verification)
- G3 (complicated trust management)
- G5 (hard to understand trust management and process of key verification)
- G6 (not set to "encrypt all" by default)
- G8 (not much guidance with trust management)

*B. Forced Updates*

Keeping a system up-to-date requires a timely use of provided security patches. However, many users are quite lax in applying security patches. Hence, nowadays, more and more software providers let not the users decide on when to patch a system but automatically apply security patches as soon as available. While this relieves the user from applying patches, it does not take into consideration the situation of the user at the moment of a forced update. The update process may require downloading a large amount of data. This is a problem when the user is temporary on a low-bandwidth connection. The update process may change security or trust relevant configuration of the application, e.g., by revoking certificates or adding new certificates that are considered trustworthy by the software provider. Often, forced updates cannot be stopped by the user, hence hinder the user.

This security mechanism is compliant with the following guidelines:
- G1 (easy to understand)
- G5 (no user decisions involved)
- G6 (keeps system up-to-date)
- G7 (does not frighten user)
- G8 (no user action necessary (or possible))
- G9 (well integrated)

This security mechanism is not compliant with the following guidelines:
- G2 (user can not decide to not apply a patch, user can not decide on time to apply patch (e.g., do not patch presentation application before presentation on CENTRIC 2012))
- G3 (in some cases user has to wait until patch was applied)

- G4 (full attention of the user when waiting for process to finish)

### C. Captchas

A captcha is a security mechanism avoiding that automated scripts use services. In theory, a captcha should be designed in a way that only humans can solve the given problem. Common captcha design requires users to read a distorted and meaningless combination of letters and numbers and enter it before he can use the service. Figure 2 shows a captcha that is even worse from a usability point of view. Another side effect of the use of captchas is that captchas may discriminate against disabled people (e.g., vision impaired people). Some websites offer different types of captchas (e.g., an image captcha and an audio captcha). Vision impaired people can decide to use the audio captcha.

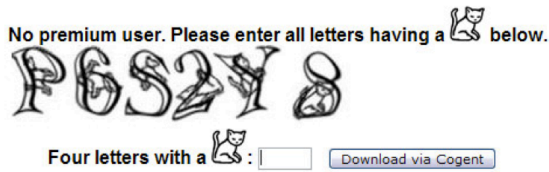

Figure 2. Complicated captcha

This security mechanism is compliant with the following guidelines:
- G5 (no user decision needed)
- G6 (always used)
- G7 (does not frighten user)
- G8 (gives instructions on how to use it)
- G1 (if multiple captchas are used, e.g., image and audio)

This security mechanism is not compliant with the following guidelines:
- G1 (if only a single image captcha is used that discriminates against disabled people)
- G2 (does not allow users to use automation tools)
- G3 (additional task without value for the user)
- G4 (unnecessary user interaction)
- G9 (many different kinds of captchas are in use)

### D. HTTPS Certificate Validation in Common Browsers

HTTPS allows for confidential and integrity protected communication on the web. For example, HTTPS is used for online banking or shopping. Nowadays HTTPS is widely used on the web. However, for a secure communication it is necessary to avoid man-in-the-middle attacks. To do so, certificates are used to authenticate the web site that one communicates with. As it is not practicable to install a certificate for each and every web site one visits, most common browsers come with preinstalled certificates of so-called Certificate Authorities (CAs). A browser accepts all certificates that have been signed by such a CA. For example, Mozilla Firefox version 14.0.1 comes with over 70 preinstalled CA certificates. The browser software developer decides on the trustworthiness of a CA (and hence on the trustworthiness of web sites), not the end user.

Figure 3 shows a typical error message of Firefox when encountering a certificate signed by an unknown CA. The text of this error message may be too complicated for average users. Above this, average users are not capable of deciding on the validity of unknown certificate anyway. As this error often occurs, the users get used to it and usually just add a security exception to the system to access the web site, bypassing the security mechanism. Adding a security exception involves multiple steps (see Figure 4 for a screenshot of the second page of the error message when clicking on "Add Exception."

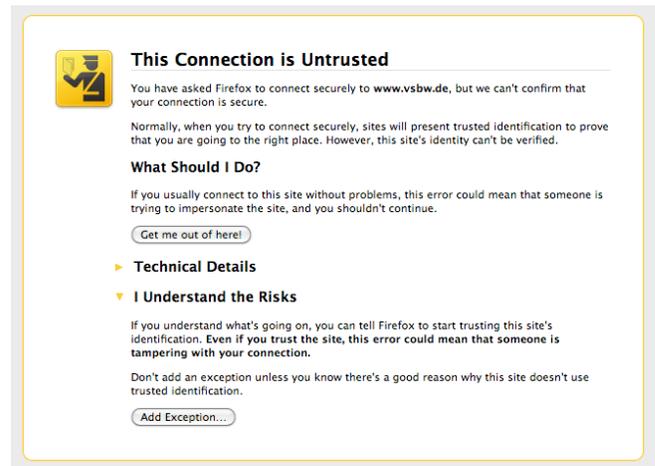

Figure 3. Typical error message of Firefox when encountering an unknown certificate

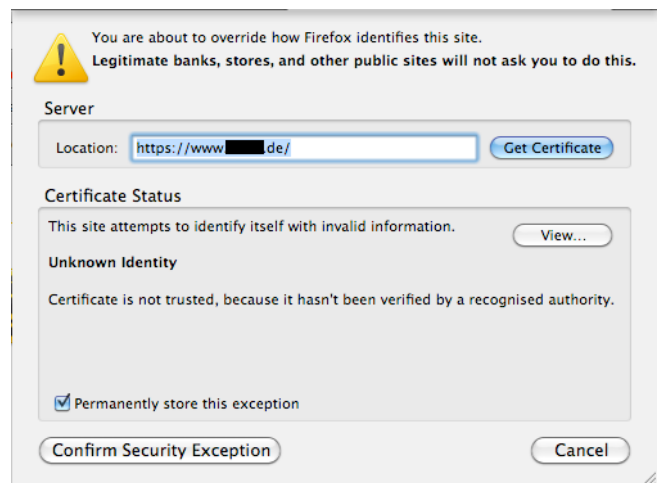

Figure 4. Second dialogue page if user clicked "Add Exception"

This security mechanism is compliant with the following guidelines:
- G6 (large number of preinstalled CAs for secure communication)

- G8 (guidance is given, however, the texts used may be not suited for average users)

This security mechanism is not compliant with the following guidelines:
- G1 (hard to understand error message given when browser encounters an unknown certificate / a certificate from an unknown CA)
- G2 (many preinstalled CA certificates, the user does not initially decide on trust relations. However, expert users can change the trust settings)
- G3 (annoying additional tasks when unknown certificate / a certificate from an unknown CA is encountered)
- G4 (error unknown certificate happens often, hence most users simply ignore the message and add a security exception)
- G5 (no informed decision possible)
- G7 (error message unknown certificate implies an ongoing attack)
- G9 (look and feel is not consistent with the rest of the browser – it changes from a website (Figure 3) to a window (Figure 4))

## V. APPLICATION OF THE PRESENTED DESIGN GUIDE: DESIGN OF AN EMAIL ENCRYPTION SOLUTION WITH GOOD USABILITY

Section IV shows usability problems of security mechanisms in common applications. This section shows for one class of application, email encryption solutions, how good usability of security mechanisms can be achieved using the guidelines presented in Section III. Section IV.A identifies the complicated key and trust management in email encryption solutions like GPGMail [5] as cause of most of the usability problems. Hence, the design of an email encryption solution presented in this paper focuses on automated key and trust management to improve the usability of email encryption.

### A. Definitions

Figure 5 shows a simplified email encryption setup: A sender wants to send an email to a receiver. The sender has his own private key ($Priv_S$) that it uses for email signatures and for decryption of emails. Please note that usually two private keys are used, one for decryption and another for signature. For the sake of simplification, this distinction is omitted in this paper. In the rest of this paper, the term "email encryption solution" is used for the presented system despite the fact that the system also decrypts and signs emails.

The sender keeps a list of email addresses and associated public keys. Public keys usually have meta information, e.g., expiration date and the like. This meta information is usually stored together with the public key and the associated email addresses in a certificate. To encrypt emails, the sender uses a public key associated with the email address of the receiver.

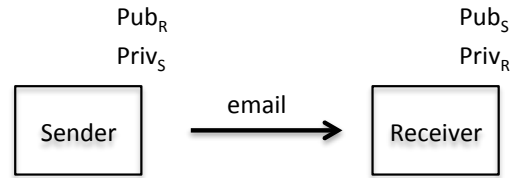

Figure 5. Keys used in email encryption

In Figure 5, the sender uses the key of the receiver, called $Pub_R$, to encrypt the email. When the receiver gets an email, he uses the public key of the sender ($Pub_S$) to verify the email signature. The receiver uses his private key ($Priv_R$) to decrypt messages encrypted with $Pub_R$. The simplified secure email setup described here is implemented by many email encryption solutions, e.g., GPG Mail [5].

### B. Approach

The email encryption solution described in this paper offers an automated key and trust management that does not require the user to take any action. Hence, it hides the most complicated part of email encryption from the user. For expert users, a manual key and trust management is still possible. The automated key and trust management is described in the following subsections in detail.

The proposed email encryption solution offers security by default: all emails are encrypted and signed by default. The necessary keys are established by the automated key and trust management if necessary and without any interaction with the user.

The user can override the default security settings: he is offered the possibility to send emails as "public postcard" by a button when composing a mail. The term "public postcard" is used as a metaphor for an unsecured email. This metaphor comes from the experience of a user, hence is a better fit than the term "unencrypted and unsigned email." The proposed email solution is realized as a plug-in to an email client, e.g., as an extension of the well-known GPGMail plugin. Existing security functionality for email is used, e.g., public key encryption and decryption as well as symmetric key encryption and decryption. The solution presented in this paper does not suppose that the receiver of an email uses the same email encryption solution. However, if the receiver of an email uses another encryption solution or no encryption solution at all, the email handling of the receiver may be a little bit more complicated then usual.

### C. Triggers for Invocation of Automated Key And Trust Management

Actions of the key and trust management are performed in the following situations:
- A user wants to send an encrypted and signed email (default) and does not have a valid public key of the receiver ($Pub_R$ missing). An automated key exchange must take place.

- A key of a receiver will expire in the near future, hence an automated rekeying is necessary.

Automated rekeying and automated key exchange are described in the following.

### D. Automated Rekeying

Automated Rekeying is invoked when the key of a receiver is about to expire. In this case, there has already been a key exchange in the past and a valid key for the receiver is still available. On the receiver side, there are two possibilities:

- The receiver has a valid key of the sender
- The receiver does not have a valid key of the sender, e.g., because the key of the sender already expired and there was no rekeying or the rekeying was not successful. This may for example be the case if the receiver does not use the same email encryption solution.

For management of keys, a list of all public keys of all past email receivers is kept. The email encryption solution regularly checks for all keys if the expiration time is near. Already expired keys are removed from the list. If the expiration time is within the time period `now+maxCheck`, a rekeying request email is sent to the owner of the associated public key. The constant `maxCheck` is a system parameter, e.g., 14 days. The rekeying request is sent by an ordinary email. It includes a certificate with the current public key of the sender and an explaining text that states something like: "Your public key with the fingerprint [fingerprint] is about to expire. Please send a new key. Please send the mail by replying to this mail and attaching a certificate with the new key." The text helps receivers that do not use the same email encryption solution to still communicate with the sender. If the receiver uses the same email encryption solution, the receiver will not see this email but a reply message will be sent. The receiver part of automated rekeying is described below. The sender waits for a reply message with a certificate holding the new public key of the receiver. The message must be signed with the old key of the receiver. If the sender does not get a reply to the rekeying request email before the expiration of the key, the key will be removed from the list. A new key exchange is necessary next time the sender sends an email to the receiver. Otherwise, it stores the received certificate and the included public key that has a validity starting in the future in the list of keys together with the current key.

If the receiver uses the email encryption solution described in this paper, each rekeying request email is deleted from the account of the user so the user never sees those requests in his emails. The following checks are performed:

- Is the email signature valid?
- Is the expiration time of the key within `now+maxCheck` (`maxCheck` is a system parameter see above)?

If the first check fails, the rekeying request is ignored to avoid denial of service attacks on public keys. If the other check fails, the key of the sender is deleted from the list. A new key exchange will be necessary in the future.

If none of the checks failed, the receiver of the key request email checks if he already has a key with a validity starting after the expiration time of the current key. If this is the case, it sends this key to the sender of the rekeying request by an encrypted and signed email that has the certificate with the new public key attached. Otherwise, the receiver of the key request email creates a new public key and associated private key with a validity starting at the expiration date of the current key and an expiration date after the starting date. The receiver creates a certificate for the public key and sends the new public key as described above. Figure 6 summarizes the control flow of the receiver on reception of a rekeying request email.

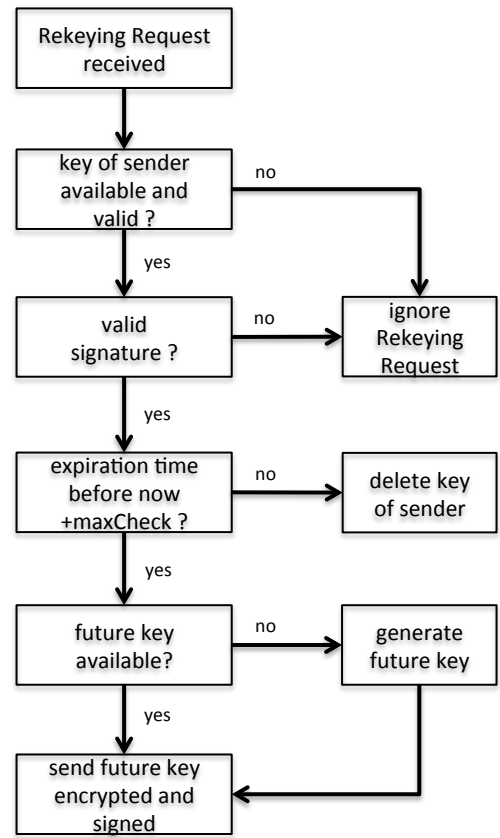

Figure 6. Control flow after receiving a rekeying request email

### E. Automated Key Exchange

The automated key exchange takes place when an email is sent to a receiver and no public key is available for this receiver. The missing key may be the result of an unsuccessful rekeying (see above). In most cases, there is no key because the sender never ever has sent an email to the receiver. In both cases, it is not clear if the receiver uses the same

email encryption solution or not. To deal with receivers not using the proposed email encryption solution, all messages used during the automated key exchange are human readable and give detailed information what must be done to interact with the sender of the email. All actions can be done manual. Please note that good usability for the receiver can be achieved if both sides use the proposed solution. Two different automated key exchange implementations are described below.

*1) Automated Key Exchange Using a Leap of Faith*

In IT security, a "leap of faith" means that at some point in time an entity has blind trust in another entity. For the automated key exchange this means that it is expected that at the time of key exchange, there is no attacker. Attackers may be present in the future. A leap of faith approach may for example protect against adversaries that hack into email accounts e.g., by guessing weak passwords. If a key exchange took place before the hacking of the email account, the exchanged keys cannot be manipulated, as the keys are stored in the email client of the user. The hacker cannot read or manipulate any email because he only has access to the account but not to the keys and all emails are encrypted and signed. However, if the key exchange takes place and the account was already hacked, a man in the middle attack is possible. Please note that automated rekeying is not prone to this attack because all messages are encrypted and signed.

Actions of the automated key exchange at sender side:
1. Generate a random string with at least 20 chars.
2. Send the random string to the receiver in an email that states: "There will be an encrypted email for you in the near future. Please use this password to decrypt the email. This email is not protected. Hence, this is the leap of faith.
3. The sender composes the key exchange message: original email is modified as follows:
   a. The sender creates an ASCII armor for the public key. An ASCII armor is a human readable representation of a public key or certificate. GPGMail offers the possibility to export certificates using an ASCII armor
   b. The sender creates an encrypted PDF that includes the original text of the message, the public key in its ASCII armor, and some explaining text: "The sender of this message wants to exchange a public key with you. Please reply with a public key in an ASCII armor in an encrypted PDF using the same password as this PDF." As password of the encrypted PDF, the sender uses the random string from step 1. The encrypted PDF is intended for receivers that do not use any email encryption solution at all.
   c. The sender creates a message encrypted by a symmetric key encryption using a string to key function to get a symmetric key from the random string generated in step 1. The message includes the original email text, some explaining text similar to the text in step 3.b and the public key of the sender.
   d. Finally, the sender prepares an email with the following text: "This message contains an encrypted email and an encrypted PDF. A password for these files was sent to you before." The encrypted PDF and the symmetrically encrypted message are both attached to the email.
4. The receiver stores the random string from step 1 in the list of keys together with an expiration time that is `leapOfFaithPeriode` in the future (`leapOfFaithPeriode` is a system value, usually a few days).

If the receiver uses the same email encryption solution, it is triggered on reception of a key exchange message. It performs the following actions:
1. Store random string together with the sender address.
2. Remove email with random string from mail server. This ensures that an attacker does not have access to the random string if the account is hacked in the future.
3. Decrypt message using the string to key function to convert the random string to a key. Retrieve public key and store it in the list of public keys.
4. Restore the original mail and encrypt it with the own public key. Delete the received mail and replace it with the mail encrypted with the public key. This avoids that an attacker can get access to the mail if the account is hacked in the future. Also, it allows forgetting the random string.
5. Compose an email including the own public key in a certificate, encrypt it with the random string using a symmetric key encryption and send it to the sender of the original message.

On reception of a reply to its key exchange email, the sender performs the following actions:
1. Retrieve random string from list of keys. If there is no random string, the message is ignored.
2. Decrypt information
   a. If it is a PDF, open it using the random key as a password. Extract the certificate and convert the ASCII armor to a binary representation of the certificate. Store the certificate in the list of keys.
   b. If it is a symmetrically encrypted message, use the string to key function to get a symmetric key from the random string.

Decrypt the message. Store the certificate in the list of keys.
3. Remove random string from the list of keys.

*2) Automated Key Exchange Using Side Channels*

The idea of the automated key exchange is to use side channels for the exchange of the random key. Compared with the leap of faith approach described in the last section, the use of a side channel improves the possibility that an attacker does not have access to the side channel used.

Side channels can be harvested from the system the email client is running on. Side channels include:

- Alternative email addresses: many people use more than one email address.
- Instant messenger addresses.
- Telephone numbers for text messaging.

The first two side channels can easily be used to send a short random string. To use side channels for key exchange, the automated leap of faith using a leap of faith is modified as follows:

In step 2, the sender sends the email message with the random string not to the same email address as the encrypted email but to a selection of available side channels for a user. If email is used as side-channel, it is very likely that the receiver collects more than one email account in the same email client. Hence, the email encryption solution has access to the side channel. An automated response is possible in this case.

*F. User Controlled Trust Management and Security as Default*

While automated key and trust management relives users from the burden of manual key and trust management, the user now does not decide on trust relations. This is against G2. The proposed usable email encryption solution lets the user decide on general trust management rules during installation. The user is presented several scenarios, which he can state that he believes in or not. These questions are used to configure the key and trust management. For example, if a user answers yes to the first question, the "leap of faith" approach is not used for key exchange.

*G. Error Handling*

Error Handling has been omitted in this section for sake of clarity of the presentation. Errors may occur during the automated key exchange or during automated rekeying. By sending a message again after a certain amount of time, the proposed email encryption solution presented in this paper deals with lost emails and the like. However, there are situations where automated key exchange or automated rekeying permanently fails, including situations in which the intended receiver of an email does not want to participate in a key exchange. As the message has already been transferred in the encrypted PDF, no further action must be taken.

## VI. EVALUATION OF THE PROPOSED EMAIL ENCRYPTION SOLUTION BASED ON THE DESIGN GUIDE

In this section, it is evaluated if the proposed email encryption solution follows the design guidelines presented in Section III:

G1 (Understandability, open for all users): the proposed solution is compliant with G1 as only good metaphors and scenarios are used for security related configurations.

G2 (Empowered users**)**: the proposed solution is compliant with G2 as the user can decide on the key and trust management configuration. Also, the user can override the security settings by sending an email without protection as "public postcard".

G3 (No jumping through hoops): the proposed solution is compliant with G3 as there are no security specific actions the user must take. It should be noted that this is not true for the receiver of an email if the receiver does not use the proposed email encryption system.

G4 (Efficient use of user attention and memorization capability): no user actions are necessary and the user does not have to memory anything, hence the proposed solution is compliant with G4.

G5 (Only informed decisions): no user actions are necessary. Hence, the proposed solution is compliant with G5.

G6 (Security as default): emails are encrypted and signed by default. Hence, the proposed solution is compliant with G6.

G7 (Fearless System): no user actions are necessary. Hence, the proposed solution is compliant with G8.

G8 (Security guidance, educating reaction on user errors): no user actions are necessary. Hence, no security guidance or education reaction on user errors is necessary.

G9 (Consistency): No user actions are necessary. Hence, there are no consistency issues. It should be noted that this is not true for the receiver of an email if the receiver does not use the proposed email encryption system.

## VII. CONCLUSION AND FUTURE WORK

This paper presented guidelines for software developers to improve the usability of security-related mechanisms. The analysis of security mechanisms in common applications showed weaknesses in the usability of security-related mechanisms as well as good examples of security usability. To demonstrate the application of the guidelines, the second part of the paper improved the identified usability weaknesses of one common application: email encryption. The approach for email encryption offers automated key and trust management to improve the usability of email encryption. The evaluation showed that the proposed email encryption solution is compliant with the usability design guide presented in this paper.

Future work will include the design of usable security mechanisms for other common problems as well as a user satisfaction study on the effectiveness of the guidelines. The guidelines presented in this paper are focused on usability for the end user. Future extensions of the design guides will focus on better usability for other groups, e.g., system administrators, testers, and developers.